\documentclass[11pt,preprint]{aastex}
\usepackage{chngpage}
\usepackage{graphicx}
\usepackage{natbib}

\shorttitle{Fall back and energy injections in GRBs}
\shortauthors{Yu \& Wu}

\begin{document}

\title{Fall back accretion and energy injections in Gamma Ray Bursts}

\author{Y. B. Yu\altaffilmark{1, 2}, X. F. Wu\altaffilmark{3, 4, 5}, Y. F. Huang\altaffilmark{1, 6},
D.M. Coward\altaffilmark{2, 7}, G. Stratta\altaffilmark{8}, B. Gendre\altaffilmark{9},
\and E.J. Howell\altaffilmark{2}}

\altaffiltext{1}{Department of Astronomy, Nanjing University, Nanjing 210093, China; hyf@nju.edu.cn}
\altaffiltext{2}{School of Physics, University of Western Australia, Crawley WA 6009, Australia;
David.Coward@uwa.edu.au}
\altaffiltext{3}{Purple Mountain Observatory, Chinese Academy of Sciences, Nanjing 210008, China;
xfwu@pmo.ac.cn}
\altaffiltext{4}{Joint Center for Particle Nuclear Physics and Cosmology of Purple Mountain
Observatory-Nanjing University, Chinese Academy of Sciences, Nanjing 210008, China}
\altaffiltext{5}{Chinese Center for Antarctic Astronomy, Chinese Academy of Sciences, Nanjing 210008, China}
\altaffiltext{6}{Key Laboratory of Modern Astronomy and Astrophysics (Nanjing University), Ministry of
Education, China}
\altaffiltext{7}{Australian Research Council Future Fellow}
\altaffiltext{8}{Universit$\grave{a}$ degli studi di Urbino `Carlo Bo', Dipartimento di Scienze di Base e Fondamenti, Piazza della Repubblica 13, I-61029, Urbino, Italy}
\altaffiltext{9}{ARTEMIS, Observatoire de la ${\rm C\hat{o}te}$ d'Azur, FRANCE}

\begin{abstract}
Intense flares that occur at late times relative to the prompt phase have been observed by
the $Swift$ satellite in the X-ray afterglows of gamma-ray bursts (GRBs). Here, we present a detailed
analysis on the fall back accretion process to explain the intense flare phase in the very early X-ray afterglow light
curves. To reproduce the afterglow at late times, we resort to the external shock by engaging energy
injections. By applying our model to GRBs 080810, 081028 and 091029, we show that their X-ray afterglow
light curves can be reproduced well. We then apply our model to the ultra-long $Swift$ GRB
111209A, which is the longest burst ever observed. The very early X-ray afterglow
of GRB 111209A showed many interesting features, such as a significant bump observed at
around 2000 s after the $Swift$/BAT trigger. We assume two constant energy injection processes in
our model. These can explain the observed plateau at X-ray wavelength in the relatively early
stage ($8.0\times10^{3}$ s) and a second X-ray plateau and optical rebrightening at
about $10^{5}$ s. Our analysis supports the scenario that a significant amount of material may
fall back toward the central engine after the prompt phase, causing an enhanced and long lived
mass accretion rate powering a Poynting-flux-dominated outflow.
\end{abstract}


\section{Introduction}
\label{sect:intro}

Gamma-ray bursts (GRBs) are short and intense unpredictable flashes of high-energy photons coming from deep
space in arbitrary directions (see Gehrels, Ramirez-Ruiz \& Fox 2009, for a review), typically lasting
for tens of seconds (Klebesadel, Strong \& Olson 1973). As the external shock model can well explain the main observed
behaviors of GRB afterglows (Rees \& ${\rm M\acute{e}sz\acute{a}ros}$ 1994; Piran 1999; Gao et al. 2013), which are widely
accepted to be from the interaction of the GRB relativistic outflow with the surrounding environment medium (${\rm M\acute{e}sz\acute{a}ros}$ \& Rees
1997a; Rhoads 1999; Sari, Piran \& Halpern 1999), it is usually recognized as the standard model. Based on duration and spectrum characteristics,
GRBs can be roughly grouped into two classes, long-soft GRBs and short-hard ones. Since the first hint for a GRB-supernovae (SN) connection
reported for GRB 980425/SN 1998bw (Galama et al. 1998), significant observational evidence has strengthened the association between
long-soft GRBs and Type Ic supernovae, which may result from the collapse of massive Wolf-Rayet (WR) stars (Bersier 2012). Since the accretion
timescale of the WR star envelope onto the central engine is consistent with the typical observed durations of long GRBs, which are tens of
seconds, it is widely accepted that long GRBs should be connected with the collapse of massive stars (Woosley 1993; ${\rm Paczy\acute{n}ski}$
1998; MacFadyen \& Woosley 1999) and WR stars are the most plausible progenitors (Chevalier \& Li 2000). Additionally, it is generally
believed that short GRBs could be due to the coalescence of two compact objects (Eichler et al. 1989; Narayan, ${\rm Paczy\acute{n}ski}$ \& Piran 1992; Gehrels et al. 2005;
Bogomazov, Lipunov \& Tutukov 2007), such as neutron star (NS) - black hole (BH) or NS-NS mergers (Janka, Ruffert \& Eberl 1998; Bloom, Sigurdsson \& Pols 1999;
Perna \& Belczynski 2002; Bethe, Brown \& Lee 2007; Tutukov \& Fedorova 2007). Before the remnant of NS-NS mergers collapse to
a BH, it is predicted that they will form an unstable millisecond pulsar (magnetar), which powers a plateau phase in the afterglow
light curve at X-ray wavelength. Recently, significant evidence obtained through observations or simulations supports this
prediction (Fan \& Xu 2006; Rowlinson et al. 2013), indicating that magnetar may be the central engine of short GRBs (${\rm Klu\acute{z}niak}$ \& Ruderman 1997).

With the development of new observational techniques and the launch of $Swift$ satellite, unexpected
features, such as multiple X-ray flares and significant optical rebrightenings, have appeared in GRB afterglows
(see Zhang 2007 for a review). For example, significant rebrightenings in optical band have been
observed from GRB 970508 (Sokolov et al. 1999), GRB 060206 (${\rm W\acute{o}zniak}$ et al. 2006), GRB 081029 (Nardini et al. 2011),
GRB 100814A (Nardini et al. 2014) and so on. As discussed in Yu \& Huang (2013), a lot of different interpretations have been
proposed to explain the optical rebrightenings. Some of these interpretations, such as the density jump model (Dai \& Lu 2002;
Lazzati et al. 2002; Dai \& Wu 2003; Tam et al. 2005), the energy injection model (Dai \& Lu 1998; Rees \& ${\rm M\acute{e}sz\acute{a}ros}$ 1998;
${\rm Bj\ddot{o}rnsson}$ et al. 2002; de Ugarte Postigo et al. 2005; Huang, Cheng \& Gao 2006; Deng, Huang \& Kong 2010; Dall'Osso et al. 2011;
Yu \& Huang 2013; Geng et al. 2013), and the microphysics variation mechanism (Kong et al. 2010), assume that the emission comes from the
same emitting region without requiring another component, which is somewhat different from the two-component jet
model (Huang et al. 2004; Liu, Wu \& Lu 2008). Note that in these mechanisms, the density jump model seems not
able to produce a significant rebrightening effectively (Huang et al. 2007).

A particular example of a burst with significant X-ray rebrightening is GRB 121027A, measured at a redshift
of $z = 1.773$ by the X-shooter spectragraph through several absorption features (Kruehler et al. 2012).
Since X-ray flares share many common features with GRB prompt emission, they are usually explained as due
to internal shocks (Burrows et al. 2005; Fan \& Wei 2005; Zhang et al. 2006), which is the well known mechanism
for prompt emission. However, GRB 121027A is so unusual that its X-ray afterglow rebrightened sharply at about
1000 s after the $Swift$/BAT trigger. The X-ray brightness increased by nearly 1000 times abruptly, and the
brightening period lasted for more than $10^{4}$ s. To explain the sharp rebrightening, Wu, Hou \& Lei (2013) proposed
a fall back accretion model based on the collapse process of massive stars for long duration GRBs. They derived a fall
back mass of $M_{\rm fb} \sim 0.9 - 2.6 M_{\odot}$, from the radius of $r_{\rm fb} \sim 3.5\times 10^{10}$ cm.
Dexter \& Kasen (2013) also pointed out that the late time power associated with the fall back material may significantly
affect the afterglow light curves.

Compared with typical GRBs, GRBs 080810, 081028 and 091029 are also very special, with significant rebrightenings existed
in the early X-ray afterglow light curve. For GRB 080810, before the shallow decay phase, two large flares appeared, interrupting
the initial sharp decay evolution. For GRB 081028, the X-ray afterglow rebrightened significantly during the sharp decay phase.
Due to the orbital gap, we can not get the complete profile of the X-ray flare of GRB 081028. Just like the above two GRBs, the
early X-ray afterglow of GRB 091029 showed similar feature. Additionally, GRB 111209A is another interesting burst. It shows a
significant bump at X-ray wavelength as well as a remarkable rebrightening in the optical band. As discussed by Stratta et al. (2013),
for GRB 111209A, internal shock models have difficulties in explaining the time lag between the
optical and high energy light curves of the flare. Interestingly, Nakauchi et al. (2013) reproduced some of
the optical/infrared and X-ray afterglow light curves reasonably well with the external shock model. In their
model, the prompt gamma-ray emissions and the afterglow emissions are attributed to a relativistic jet, while
the superluminous-supernova like bump comes from a non-relativistic cocoon fireball. However, their model has
difficulty in reproducing some of the X-ray plateaus and the optical rebrightenings observed at late stages.
Energy injections from late activities of the central engines seem to be a natural explanation for the
rebrightening of many GRB afterglows. In this paper, we will use the fall back accretion model
to interpret the unusual afterglow light curves of the above four GRBs.

The structure of our paper is as follows. The observations are summarized in Section 2. In Section 3, we introduce the
fall back accretion model and energy injection process in the external shock. We reproduce the unusual afterglow light
curves of the above four GRBs and present our numerical results in Section 4. Our results show that the observed X-ray bumps
and other features can be well reproduced. Finally, we conclude in Section 5 with a discussion and summary. We use the following
cosmological parameters, $H_{0}$ = 71 km ${\rm s^{-1}~Mpc^{-1}}$, $\Omega_{\Lambda}$ = 0.73 and $ \Omega_{M} $ = 0.27, throughout
the paper. A list of important symbols in this paper is summarized in Table 1.

\begin{deluxetable}{ccccc}
\tabletypesize{\scriptsize}
\tablewidth{0pt}
\tablecaption{List of symbols.\label{TABLE:Fit1}}

\tablehead{
        \colhead{Symbol} &
        \colhead{Description}}
\startdata
$\eta$            & Efficiency of converting Blandford-Znajek power to the X-ray radiation \\
$f_{\rm b}$       & Beaming factor of the jet \\
$M_{\bullet}$     & Mass of the central black hole \\
$a_{\bullet}$     & Spin of the central black hole \\
$\dot{M_{\rm p}}$ & Peak accretion rate \\
$s$               & Sharpness of the peak \\
$\theta$          & Half opening angle \\
$\epsilon_{\rm e}$ & Fraction of electron energy \\
$\epsilon_{\rm B}$ & Fraction of magnetic energy \\
$M_{\rm ej}$      & Initial ejecta mass \\
$E_{\rm iso}$     & Isotropic energy \\
$p$               & Index of the power-law distribution of the electrons \\
$t_{\rm start}$   & Beginning time of the energy injection \\
$t_{\rm end}$     & Ending time of the energy injection \\
$Q$               & Energy injection rate \\
\enddata
\end{deluxetable}

\section{Sample}
\label{sect:obs}

\subsection{GRB 080810}

Both $Swift$ and $Fermi$ were triggered by GRB 080810 at 13:10:12 UT on August 10 2008 and the position determined from the UVOT/$Swift$ analysis is RA(J2000) = $23^{h}47^{m}10.48^{s}$, Dec(J2000) = $+00^{\circ}19^{'}11.3^{''}$ (Holland \& Page 2008), which is consistent with the localizations of Robotic Optical Transient Search Experiment (Rykoff 2008) and Nordic Optical Telescope (NOT) (de Ugarte Postigo et al. 2008). The $T_{90}$ measured in the 15 -- 150 keV band by BAT/$Swift$ is $108 \pm 5$ s. The redshift of GRB 080810 determined by Keck/HIRES is $z = 3.355 \pm 0.005$ (Prochaska et al. 2008), which was confirmed by NOT (de Ugarte Postigo et al. 2008) and RTT150 (Burenin et al. 2008). XRT/$Swift$ followed GRB 080810 in Window Timing (WT) mode throughout the first orbit and started to observe it in Photon Counting (PC) mode from the beginning of the second orbit. The X-ray afterglow light curve of GRB 080810 is very strange, with two large flares in the early stage, which can be well fitted with Fast Rise Exponential Decay (FRED) profiles (Page et al. 2009). After the flares, the X-ray afterglow evolved into a shallow plateau phase, which is followed by a normal decay phase. During the shallow plateau phase, there is no way to constrain the start time of the plateau due to orbital gaps in the data. The observed X-ray afterglow light curve of GRB 080810 is plotted
in Fig 1.

\subsection{GRB 081028}

The BAT/$Swift$ was triggered by GRB 081028 at 00:25:00 UT on 28 October 2008 (Guidorzi et al. 2008). The X-ray Telescope onboard $Swift$ satellite started to observe GRB 081028 191 s after the trigger and discovered a bright and fading X-ray afterglow. The UVOT/$Swift$ began to follow GRB 081028 210 s after the trigger and localized it at the position of RA(J2000) = $08^{h}07^{m}34.76^{s}$, Dec(J2000) = $+02^{\circ}18^{'}29.8^{''}$ with an estimated error of 1.5 arcsec (Evans et al. 2008). The redshift measured from several absorption features including FeII and SII is $z = 3.038$ (Berger et al. 2008). The rest frame isotropic energy and peak energy measured in the $1 - 10^{4}$ keV energy range are $E_{\rm iso} = (1.1 \pm 0.1) \times 10^{53}$ erg and $E_{\rm p} = 222$ keV respectively (Margutti et al. 2010), which are consistent with the Amati relation (Amati 2006). There is a flat phase in the early X-ray afterglow light curve, followed by a steep decay phase with flares, which dominated the afterglow light curve from about 690 s. After the orbital gap between the peak of the flare and the end of the sharp decay phase, a significant rebrightening appears in the X-ray afterglow light curve. Interestingly, there is no evidence for spectral evolution during the rebrightening, after which the X-ray afterglow light curve entered the normal decay phase. The observed X-ray afterglow light curve of GRB 081028 is plotted in Fig 1.

\subsection{GRB 091029}

GRB 091029 triggered the $Swift$/BAT at 03:53:22 UT on 29 October 2009 and the $T_{90}$ measured in the 15 -- 350 keV band is $39.2 \pm 5$ s (Barthelmy et al. 2009).
Given the fluence of $(2.4 \pm 0.1) \times 10^{-6}$ erg~${\rm cm^{-2}}$ measured in the 15 -- 150 keV energy range and the redshift of $z = 2.752$ (Chornock, Perley \& Cobb 2009), the isotropic energy released in the rest frame of GRB 091029 is $E_{\rm iso} = 8.3 \times 10^{52}$ erg. The X-ray afterglow light curve of GRB 091029 shows an initial steep
decay and the decay index is $3.69 \pm 0.10$, which is consistent with being the high latitude emission of the prompt phase (Fenimore, Madras \& Nayakshin 1996; Kumar \& Panaitescu 2000).
After the sharp decay, there is a steep X-ray bump, which interrupts the smooth early-time temporal evolution. The X-ray afterglow light curve after 700 s can be fitted with
a broken power law and the best-fitting decay indices are $-0.12 \pm 0.10$ and $1.20 \pm 0.04$ with a break time of $t_{\rm b} = 7.4 \pm 1.8$ ks (Filgas et al. 2012).
Interestingly, the X-ray spectral index of GRB 091029 remains constant throughout the observations. The observed X-ray afterglow light curve of GRB 091029 is plotted in Fig 1.

\subsection{GRB 111209A}

GRB 111209A was detected at 07:12:08 UT on Dec 09 2011 by $Swift$ and was located at the position of RA(J2000) = $00^{h}57^{m}22.63^{s}$,
Dec(J2000) = $-46^{\circ}48^{'}03.8^{''}$ (Hoversten et al. 2011). The 1-sec peak photon flux of GRB 111209A measured by the $Swift$/BAT in
the 15 -- 150 ${\rm keV}$ band was $0.5 \pm 0.1 {\rm ~ph~cm^{-2}~s^{-1}}$ with the intrinsic peak of the spectrum $E_{\rm p} = 520 \pm 89$ keV.
Assuming a simple power law, the best fit spectral index of the X-ray spectrum at about 5 days is $\beta_{X} = 1.8 \pm 0.4$ (Stratta et al. 2013),
which is consistent with the prediction of the external shock model for fast cooling electrons (Sari, Piran \& Narayan 1998). The redshift measured by identifying
several absorption lines in the host galaxy of GRB 111209A with the X-shooter spectrograph mounted at the Kueyen unit of the VLT on Cerro Paranal was $z = 0.677$
(Vreeswijk, Fynbo \& Melandri 2011). The luminosity distance of GRB 111209A is then 4.1 ${\rm Gpc}$. Follow-up observations from the ground-based
telescopes generated the multi-band afterglow light curves of GRB 111209A, which is a special burst with a
significant X-ray bump observed at about 2000 s after the $Swift$/BAT trigger.

The $Swift$/XRT began to observe GRB 111209A 425 s after the $Swift$/BAT trigger (Hoversten et al.
2013), revealing a bright afterglow. The very early X-ray afterglow light curve shows an initial shallow decay phase lasting for about 2000 s and the best fit decay index is $\alpha = 0.544 \pm 0.003$ (Gendre et al. 2013). What makes the X-ray afterglow light curve unusual is the significant bump beginning at about 2000 s after the $Swift$/BAT trigger (see Fig 2). At the end of the shallow decay phase, the X-ray afterglow light curve evolved into the so called steep decay phase, which corresponds to the true end of the prompt phase and is usually explained as the tail emission coming from high latitude (Fenimore et al. 1996; Kumar \& Panaitescu 2000). The steep decay of the afterglow light curve at X-ray wavelength was best fit by a simple power law ($f_{\nu}$ = $t^{-\alpha}$) with an index of $\alpha = 4.9 \pm 0.2$ (Gendre et al. 2013). The X-ray afterglow light curve after the steep decay phase can be described by a broken power law function, a plateau with decay index of $0.5 \pm 0.2$ and a normal decay with an index of $1.51 \pm 0.08$ (Gendre et al. 2013). According to Gendre et al. (2013), a thermal component appeared in the prompt spectrum of GRB 111209A at the beginning of the $Swift$/XRT observation, but it disappeared very soon.

The optical afterglow of GRB 111209A was detected by the $Swift$/UVOT in all seven filters 427 s after the $Swift$/BAT trigger. In the early optical afterglow light curve, there are some flares, which may come from the contamination of the prompt emission. The optical afterglow light curve at late times shows unusual behavior with a marked rebrightening at about $10^{5}$ s after the $Swift$/BAT trigger, and is quite different from the typical optical afterglows. In R band, the afterglow brightness increased from $R\sim20.9$ magnitude to a peak value of $R\sim20.0$ magnitude (see Fig 3), implying a possible supply of a large amount of energy at such late stage.

\section{Model}
\label{sect:model}

\subsection{Fall back accretion model}

Kumar, Narayan \& Johnson (2008b) associated the early X-ray light curves of GRBs with the fall back accretion of
the stellar envelope from a massive star and provided some observational consequences.
Wu et al. (2013) proposed a fall back accretion model to explain the step-like rebrightening of the
X-ray afterglow of GRB 121027A observed at around $10^{3}$ s after the $Swift$/BAT trigger with a duration of more
than $10^{4}$ s. Wu et al. (2013) assumed that, in the fall back accretion model, the central engine is a BH. The Blandford-Znajek (BZ)
mechanism powers GRB jet by extracting magnetic energy  from the rotating BH. A large proportion of the stellar envelope that have survived
mass loss will undergo fall back during the last stage of evolution of the progenitor. The accretion process of the helium envelope may significantly
increase the X-ray brightness observed in some GRBs. In this paper, we will use the fall back process to explain the very early X-ray afterglow light
curves of the above GRBs. For completeness, we first describe the fall back accretion model briefly (for details, see Wu et al. 2013).

According to Wu et al. (2013), the evolution of mass ($M_{\bullet}$), angular momentum ($J_{\bullet}$) and spin (${a_{\bullet}}$) of the BH are described as
\begin{equation}
\frac{dM_{\bullet}c^{2}}{dt}=\dot{M}c^{2}E_{\rm ms} - \dot{E}_{B},
\end{equation}
\begin{equation}
\frac{dJ_{\bullet}}{dt}=\dot{M}L_{\rm ms} - T_{B},
\end{equation}
\begin{equation}
\frac{da_{\bullet}}{dt}=(\dot{M}L_{\rm ms} - T_{B})c/(GM_{\bullet}^{2})
- 2a_{\bullet}(\dot{M}c^{2}E_{\rm ms} - \dot{E}_{B})/(M_{\bullet}c^2),
\end{equation}
where $\dot{M}$, $\dot{E_{B}}$ and $T_{B}$ are the fall back accretion rate, BZ jet power and magnetic
torque in the BZ process, respectively. According to Novikov \& Thorne (1973), $E_{\rm ms}$ and $L_{\rm ms}$
are the specific energy and angular momentum respectively, which are related to the radius of the marginally
stable orbit (Bardeen, Press \& Teukolsky 1972).

The initial accretion rate in the fall back process increases with time as $\dot{M} \propto t^{1/2}$
(MacFadyen, Woosley \& Heger 2001; Zhang, Woosley \& Heger 2008; Dai \& Liu 2012), and the late-time accretion
behavior follows $\dot{M} \propto t^{-5/3}$ (Chevalier 1989; Dexter \& Kasen 2013) with a break time $t_{\rm p}$.
Therefore, as assumed by Wu et al. (2013), we take the accretion rate of the fall back as a smooth-broken-power-law function
\begin{equation}
\dot{M} = \dot{M_{\rm p}}\left[\frac{1}{2}\left(\frac{t-t_{0}}{t_{\rm p}-t_{0}}\right)^{-\alpha_{r}s}
+ \frac{1}{2}\left(\frac{t-t_{0}}{t_{\rm p}-t_{0}}\right)^{-\alpha_{d}s}\right]^{-1/s},
\end{equation}
where $\alpha_{\rm r} = 1/2$, $\alpha_{\rm d} = -5/3$ and $s$ describes the sharpness of
the peak. $\dot{M_{\rm p}}$ is the peak accretion rate and $t_{0}$ is the start time of fall back.
The BZ jet power from a BH with angular momentum $J_{\bullet}$ and mass $M_{\bullet}$ is
(Lee, Wijers \& Brown 2000; Wang, Xiao \& Lei 2002; Mckinney 2005; Lei \& Zhang 2011)
\begin{equation}
\dot{E_{B}} = 1.7 \times 10^{50}a_{\bullet}^{2}(M_{\bullet}/M_{\odot})^{2}B_{\bullet,15}^{2}F(a_{\bullet})~\rm
erg~\rm s^{-1},
\end{equation}
where $F(a_{\bullet})$ is a function of spin of the BH (Wu et al. 2013).

\subsection{External shock}

In our study, we use the equations for beamed GRB outflows developed by Huang, Dai \& Lu (1999) and
Huang et al. (2000) to calculate the general dynamical evolution of the external shock. These equations
are concise and the calculations can be easily extended to the deep Newtonian phase (Huang \& Cheng 2003).
In our calculations, the effects of electron cooling, lateral expansion, and equal arrival time surfaces were
all incorporated. However, note that if some subtle effects such as the adiabatic pressure and radiative losses
were considered, then the dynamics could be slightly different (van Eerten et al 2010; Pe'er 2012; Nava et al. 2013).

Various forms of energy injection in GRB afterglows have been studied previously by many authors.
For example, Dai \& Lu (1998) suggested that a new-born magnetar will lose its rotation energy through
dipole radiation. If this kind of energy is injected into the external shock, then the injection power
will take the form of $dE_{\rm inj}/dt \propto (1+t/T)^{-2}$, where $t$ is the time in the rest frame
and $T$ is the spin-down timescale. In this case, the injection power will roughly be a constant when $t \ll T$,
but will decrease quickly as $dE_{\rm inj}/dt \propto t^{-2}$ when $t \gg T$. Zhang \& ${\rm M\acute{e}sz\acute{a}ros}$ (2001b)
generally assumed a power-law function for the injection luminosity, $dE_{\rm inj}/dt \propto t^{q}$.
They argued that $q = 0$ (i.e., a constant injection power) could correspond to
many realistic situations. According to Zhang \& ${\rm M\acute{e}sz\acute{a}ros}$ (2001a),
for many kinds of central engines which involves a black hole plus a relatively long-lived
accretion disk system, the timescale of energy injection to the external shock may be much
longer than that of the prompt burst. Following their argument, as suggested by Kong \& Huang (2010),
Yu \& Huang (2013) then assumed the energy injection power as $dE_{\rm inj}/dt = Q t^{q} (t_{\rm start}<t<t_{\rm end}) $ to
interpret the unusual features of the observed multi-band afterglow light curves of GRB 081029.
Here, $Q$ and $q$ are constants, while $t_{\rm start}$ and $t_{\rm end}$ are the beginning and ending time
of the energy injection respectively.

In our current study, the energy injection should mainly be from the fall back process. But to account for some special
features observed at late times, we further need to assume other periods with constant injection power. Considering all
these energy injections, we then use the following equation to calculate the evolution of the bulk Lorentz factor of the external shock (Kong \& Huang 2010)
\begin{equation}
\frac{d\gamma}{dt}=\frac{1}{M_{ej}+\epsilon m+2(1-\epsilon) \gamma
m}\times\left(\frac{1}{c^{2}} \frac{dE_{inj}}{dt}-(\gamma^{2}-1) \frac{dm}{dt}\right),
\end{equation}
where $m$ is the swept-up interstellar medium (ISM) mass, $M_{ej}$ is the initial ejecta mass,
and $\epsilon$ is the radiative efficiency.

\section{Numerical Results}
\label{sect:nume}

\subsection{Parameters}

In this study the parameter of the beaming factor is taken as a constant of $f_{\rm b} = 0.01$,
corresponding to a half opening angle of $\theta = 0.1$ radian, which is a typical value for the relativistic outflow of GRB.
Given the fluence and redshift of the GRBs, we can calculate the isotropic energy released in the rest frame. By assuming the isotropic
energy as the isotropic-equivalent kinetic energy, we can obtain a measure of the initial ejecta mass with a reasonable Lorentz factor.
The start time of the energy injection depends on the start time of the rebrightening or plateau of the observed afterglow emission.
The end time of the energy injection is related to the peak time of the rebrightening or the end time of the plateau. The peak accretion
rate is estimated from the amplitude of the rebrightening. Actually, the peak accretion rate is coupled with the spin of the central BH, which
can be seen from the equation (5). When fitting the X-ray afterglow emission of the above four GRBs, we also require a measure of the efficiency of
converting the BZ power to the X-ray radiation. As the efficiency depends on a number of effects, such as radiative losses of the outflow, the mass,
spin, and angular momentum of the central BH etc, in our numerical calculations, we assume the efficiency as $\eta = 0.01$ for simplicity.
Additionally, we use typical values for long GRBs for the fraction of magnetic energy, the fraction of electron energy, and the index of the
power-law distribution of the shocked electrons.

\subsection{GRBs 080810, 081028 and 091029}

We apply our model to these three GRBs and Figure 1 illustrates our numerical results. The X-ray bumps observed
in the early stage come from the fall back accretion process and the late X-ray plateau or rebrightening can be
explained as due to energy injection from the central engine. It is shown that our model can satisfactorily reproduce
the observed X-ray afterglow light curves. For GRB 080810, we assume the peak accretion rate as
$\dot{M_{\rm p}} = 4.0 \times 10^{-5}M_{\odot}~\rm s^{-1}$ to interpret the early X-ray bump and the late X-ray plateau observed
at about $10^{3}$ s can be fitted well with a constant energy injection rate of $Q$ = $5.0\times10^{47}$ ${\rm erg}$ ${\rm s^{-1}}$.
Note that there is no way to constrain the start time of the X-ray plateau, we just assume $t_{\rm start}$ to be 600 s. To explain
the observed unusual X-ray afterglow features of GRB 081028, the peak accretion rate and the constant energy injection rate are
taken as $\dot{M_{\rm p}} = 4.0 \times 10^{-4}M_{\odot}~\rm s^{-1}$ and $Q$ = $5.0\times10^{47}$ ${\rm erg}$ ${\rm s^{-1}}$ respectively.
Due to orbital gaps in the data, the sharpness of the peak can not be constrained tightly. The early X-ray bump and the late X-ray
plateau of GRB 091029 can also be satisfactorily reproduced by assuming the peak accretion rate and the energy injection rate to be
$\dot{M_{\rm p}} = 3.0 \times 10^{-5}M_{\odot}~\rm s^{-1}$ and $Q$ = $8.0\times10^{47}$ ${\rm erg}$ ${\rm s^{-1}}$ respectively. The best
fit parameters for the observed X-ray afterglow light curves of these three GRBs during the fall back accretion and external shock
process are summarized in Table 2.

\subsection{GRB 111209A early afterglow}

The X-ray bump of GRB 111209A appears at about 2000 s, which corresponds to a rest frame duration of
$t_{0} \sim 1200$ s. From this timescale, we can derive the minimum fall back radius of matter as
\begin{equation}
r_{\rm fb} \simeq 7.7 \times 10^{10} (M_{\bullet}/3M_{\odot})^{1/3}(t_{0}/1200~\rm s)^{2/3}~\rm cm.
\end{equation}
The rise of the bump lasts for about 1100 s, which corresponds to
$t_{\rm p} - t_{0} \sim 1100/(1+z)~\rm s \sim 650$ s. According to this timescale, the total mass accreted
during the fall back process should be
\begin{equation}
M_{\rm fb} \simeq 2\dot{M_{\rm p}}(t_{\rm p} - t_{0})/3 \simeq 1.6 \times 10^{-2}L_{\rm
X,iso,50}a_{\bullet}^{-2}X^{-1}(a_{\bullet})\eta_{-2}^{-1}f_{\rm b,-2}M_{\odot},
\end{equation}
where $L_{\rm X,iso,50} = L_{\rm X,iso}/10^{50}$ and $X(a_{\bullet})$ is a function of spin of the BH
(Wu et al. 2013).

We obtain the time evolution of the BZ power by carrying out numerical calculations of Eqs.(1) - (5),
and compare the results with the observed X-ray bump in GRB 111209A. A complete monitoring before
and after the flare peak epoch at about 2000 s after the $Swift$/BAT trigger was available only with the TAROT
R-band and Konus-Wind (KW) data. Assuming a Gaussian function, Stratta et al. (2013) measured a peak epoch
at $(2460 \pm 50)$ s after the $Swift$/BAT trigger with a width of $\sim$ 130 s using the TAROT data, while a
peak epoch at $(2050 \pm 10)$ s with a width consistent with the optical one using the KW data. Because there
is no observations in X-ray band during the bump, we assume the peak time as $t_{\rm p} \sim 2050/(1+z)$ s.

In our calculations, we take the sharpness of the peak as $s = 1.9$. The mass and spin of the BH are initially
set as $M_{\bullet} = 3M_{\odot}$ and $a_{\bullet} = 0.9$ respectively, as assumed by Wu et al. (2013).
Considering the luminosity at X-ray wavelength in the prompt phase, we assume the peak accretion rate as
$\dot{M_{\rm p}} = 2.0 \times 10^{-4}M_{\odot}~\rm s^{-1}$. To explain the plateau observed at X-ray wavelength in
the early stage, we assumed a constant energy injection rate of $Q = 9.0\times10^{47}$ ${\rm erg}$ ${\rm
s^{-1}}$ with the start and the end time of the energy injection as $t_{\rm start} = 8.0 \times 10^{3}$ s,
$t_{\rm end} = 1.6 \times 10^{4}$ s respectively. The steep decay phase observed at X-ray wavelength starting
at around 20000 s is the end of the prompt phase and can be explained as the high latitude emission of the
prompt phase (Fenimore et al. 1996; Kumar \& Panaitescu 2000), whose theoretical decay index is $2 + \beta_{\rm X}$, where
$\beta_{\rm X}$ is the spectral index at X-ray wavelength. Figure 2 illustrates our numerical fit to the observed
X-ray bump in GRB 111209A: our calculations start at $t_{0} = 2000/(1+z)$ s.

\subsection{GRB 111209A late afterglow}

Assuming the relativistic outflow expands in a uniform ISM, we calculate the dynamical evolution of the ejecta
numerically based on the external shock by adding an energy injection process, and reproduce the observed multi-band
afterglow light curves at late times. The fluence of GRB 111209A in the 15 -- 150 ${\rm keV}$ band measured by
$Swift$/BAT is $(360 \pm 10) \times 10^{-7}$ ${\rm erg~cm^{-2}}$, indicating that the isotropic energy released
in the rest-frame should be $E_{0,iso}$ = $4.3\times10^{52}$ ${\rm erg}$. In our calculations, other parameters
have been evaluated typically, such as the initial ejecta mass $M_{ej}$ = $4.0\times10^{-4}M_{\odot}$,
the fraction of magnetic energy $\epsilon_{B}$ = 0.004, the fraction of electron energy $\epsilon_{e}$ = 0.04,
and the index characterizing the power-law distribution of the shocked electrons $p=2.25$.

Possible theoretical interpretations about the remarkable optical rebrightening observed at about $10^{5}$ s
after the $Swift$/BAT trigger were discussed in Stratta et al. (2013). In this paper, we assume another
energy injection imposed on the external shock with $Q$ = $6.0\times10^{46}$ ${\rm erg}$ ${\rm s^{-1}}$,
$q$ = 0, $t_{\rm start}$ = $6.5\times10^{4}$ s, and $t_{\rm end} = 8.1 \times10^{4}$ s to interpret this
significant rebrightening. The total energy injection during this phase is about $E_{\rm inj}$ = 2.3 $E_{0}$,
where $E_{0}$ = $(1-\cos\theta)E_{0,\rm iso}$ is the collimation-corrected energy.
Zhang \& ${\rm M\acute{e}sz\acute{a}ros}$ (2002) pointed out that, the brightness of the afterglow light curve
will be enhanced significantly if the injected energy was higher than the original kinetic energy of the ejecta.
It should be noted that, if the energy injection form was kinetic-energy-dominated but not Poynting-flux-dominated
(Usov 1994; ${\rm M\acute{e}sz\acute{a}ros}$ \& Rees 1997b), the reverse shock produced during the injection
phase will also generate an obvious rebrightening in the afterglow light curve (Rees \& ${\rm M\acute{e}sz\acute{a}ros}$ 1998; Kumar \& Piran 2000).
In our model, the energy injection was generated by the central engine through the BZ mechanism (i.e., Poynting-flux-dominated),
in which case, there will be no reverse shock generated during the injection process.

Our model can satisfactorily reproduce the observed multi-band afterglow light curves of GRB 111209A at late times with the external shock mechanism
and parameters described above. By superposing the theoretical luminosity in the X-ray band from the fall back process in the early stage and
the external shock at late times, we can fit all the observational data at X-ray wavelength. The observed X-ray and optical afterglow light curves
of GRB 111209A and our best theoretical fit are illustrated in Figures 2 and 3 respectively.

\begin{figure}
   \begin{center}
   \plotone{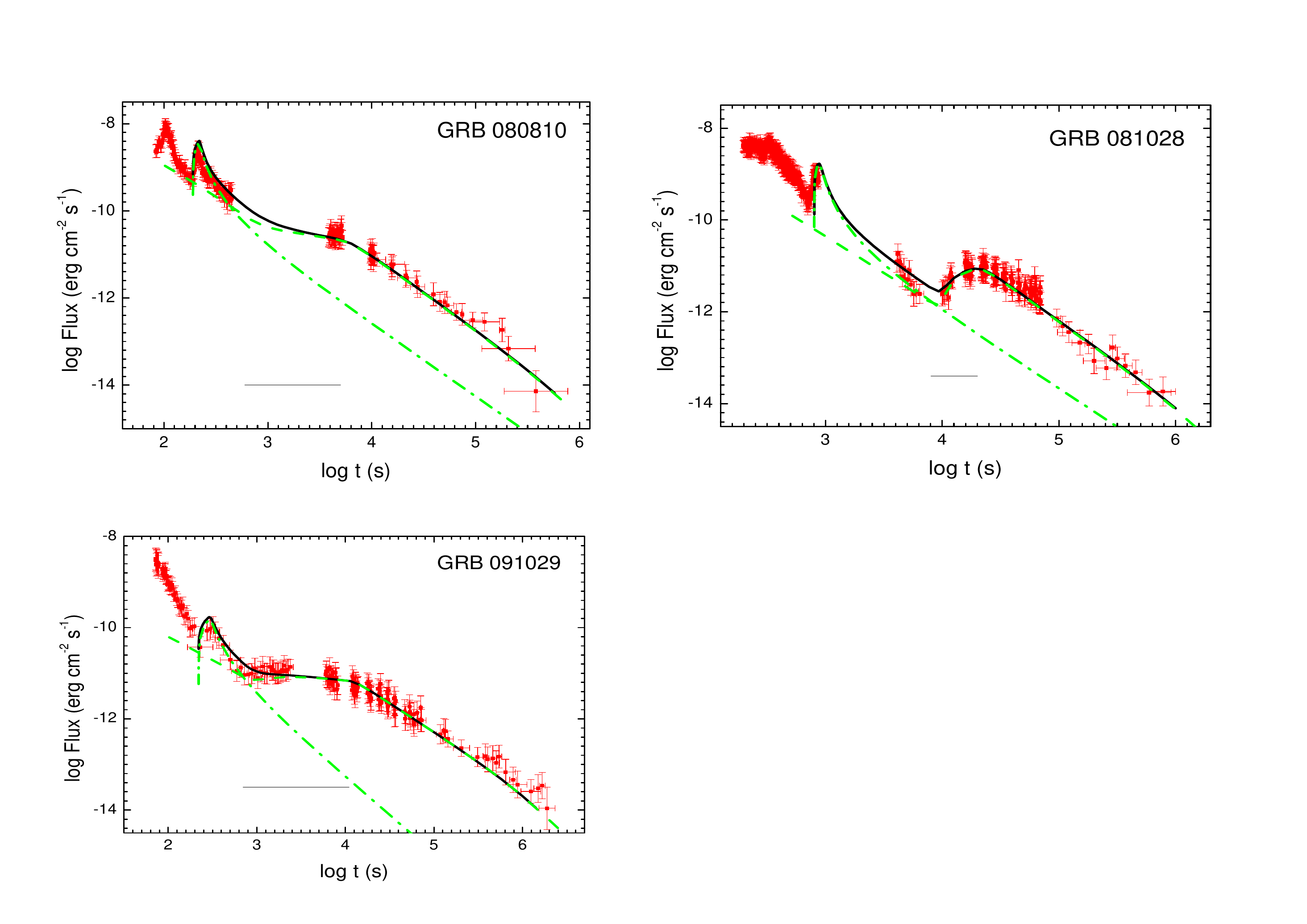}
   \caption{ Observed X-ray afterglow light curves of GRBs 080810, 081028 and 091029 and our best fit by using the fall back accretion model.
The solid points are data from the $Swift$ XRT data repositories; the dash-dotted and dashed lines correspond to the theoretical fluxes
in the X-ray band from fall back and external shock processes respectively. The solid line is the combined light curve obtained from our
theoretical model. The horizontal line at the bottom of each panel represents the duration of the energy injection phase.}
   \label{Fig:plot1}
   \end{center}
\end{figure}

\begin{figure}
   \begin{center}
   \plotone{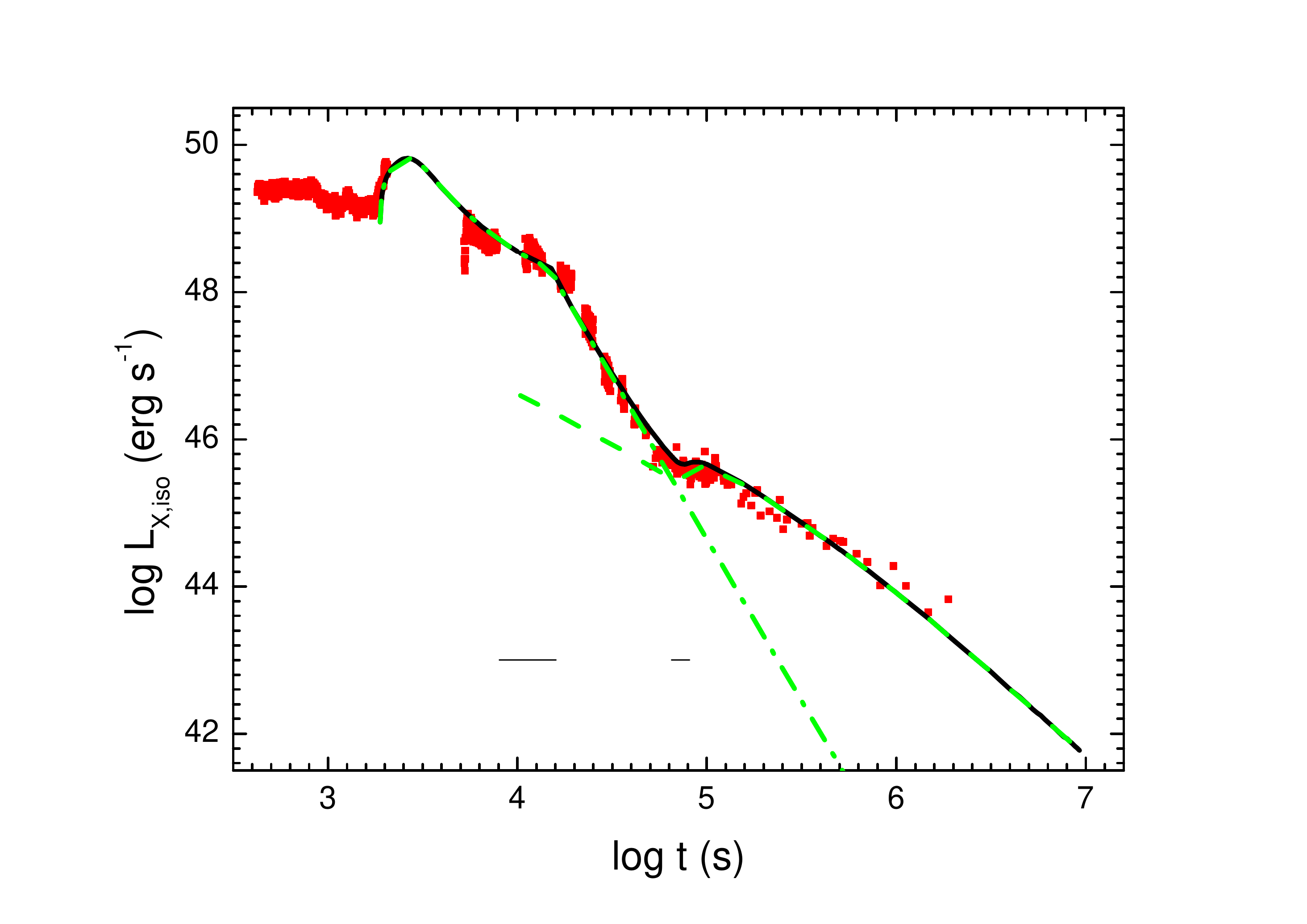}
   \caption{ Observed X-ray afterglow light curve of the ultra-long GRB 111209A and our best fit by using the fall back accretion model.
The solid points represent the observed data from Stratta et al. (2013); the dash-dotted and dashed lines correspond to the theoretical
luminosity in X-ray band from fall back and external shock processes respectively. The solid line is the overall light curve for GRB 111209A
from our theoretical model. The two straight lines at the bottom represent the periods of the two energy injection phases.}
   \label{Fig:plot2}
   \end{center}
\end{figure}

\begin{figure}
   \begin{center}
   \plotone{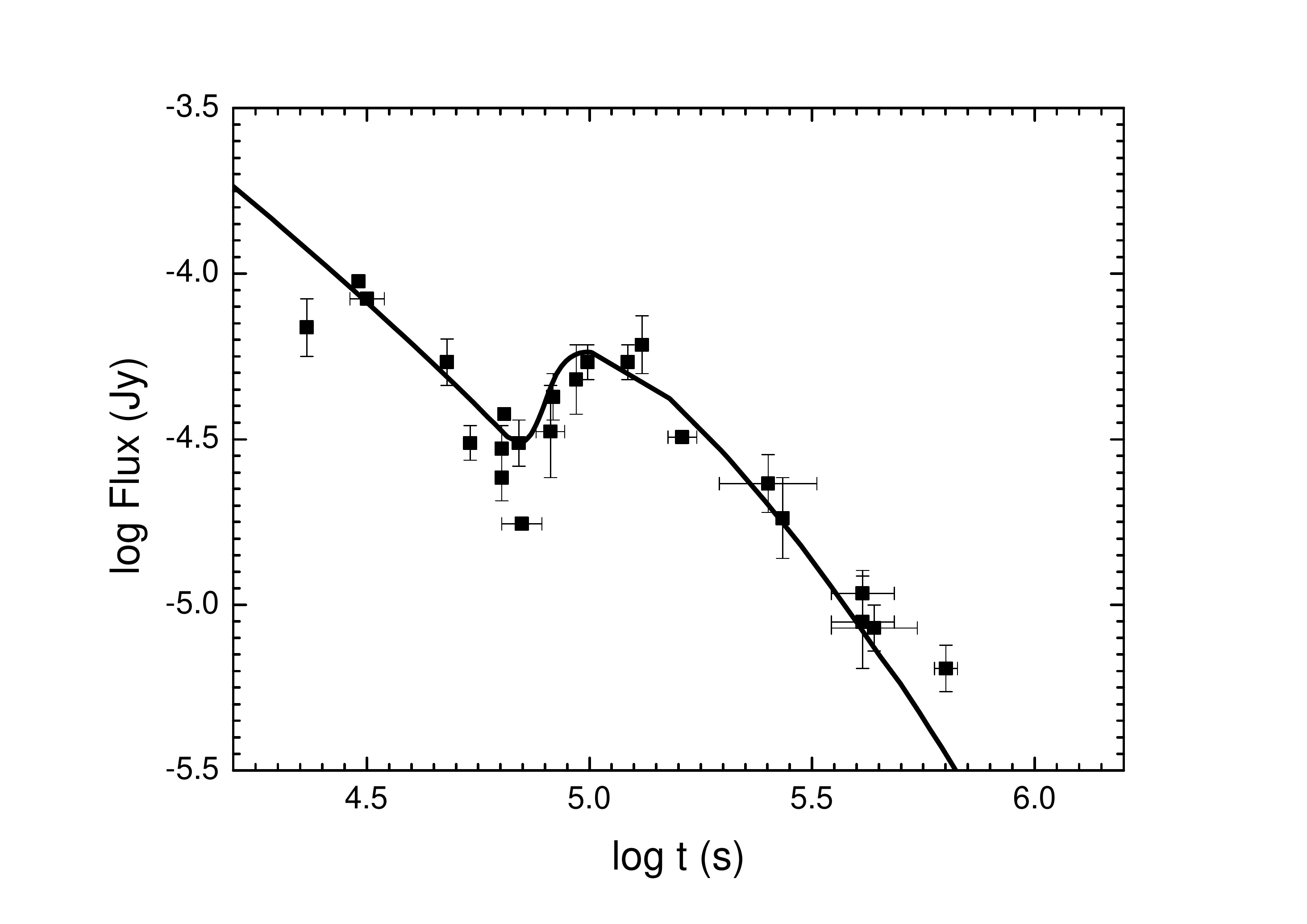}
   \caption{Observed R-band optical afterglow of the ultra-long GRB 111209A and our best fit by using the same model as in Figure 2.
The observational data are taken from Stratta et al. (2013). The solid line is our theoretical afterglow light curve in optical band corrected for extinction.}
   \label{Fig:plot3}
   \end{center}
\end{figure}

\begin{deluxetable}{ccccc}
\tabletypesize{\scriptsize}
\tablewidth{0pt}
\tablecaption{Best fit parameters for the X-ray afterglow light curves of selected GRBs during the fall back accretion and external shock process.\label{TABLE:Fit2}}

\tablehead{
        \colhead{Parameter} &
        \colhead{GRB 080810} &
        \colhead{GRB 081028} &
        \colhead{GRB 091029} &
        \colhead{GRB 111209A}}
\startdata
$\eta$            & 0.01           & 0.01           & 0.01     & 0.01        \\
$f_{\rm b}$       & 0.01           & 0.01           & 0.01     & 0.01        \\
$M_{\bullet}$ ($M_{\odot}$)        & 3.0            & 3.0      & 3.0  & 3.0  \\
$a_{\bullet}$     & 0.6            & 0.9            & 0.9      & 0.9         \\
$\dot{M_{\rm p}}$ ($M_{\odot}/s$)  & $4.0 \times 10^{-5}$      & $4.0 \times 10^{-4}$
                  & $3.0 \times 10^{-5}$            & $2.0 \times 10^{-4}$   \\
$s$               & 6.0            & 1.9            & 6.0      & 1.9         \\
$\theta$ (rad)     & 0.1           & 0.1            & 0.1      & 0.1         \\
$\epsilon_{\rm e}$ & 0.1           & 0.1            & 0.05     & 0.04        \\
$\epsilon_{\rm B}$ & 0.01          & 0.01           & 0.005    & 0.004       \\
$M_{\rm ej}$ ($M_{\odot}/s$)       & $1.8 \times 10^{-4}$      & $1.8 \times 10^{-4}$
                   & $1.8 \times 10^{-4}$  & $4.0 \times 10^{-4}$            \\
$E_{\rm iso}$ (erg)& $8.0 \times 10^{52}$           & $8.0 \times 10^{52}$
                   & $8.0 \times 10^{52}$           & $4.3 \times 10^{52}$   \\
$p$                & 2.5           & 2.5            & 2.1      & 2.25        \\
$t_{\rm start,1}$ (s)  & 600  & 8000 & 700 & 8000               \\
$t_{\rm end,1}$ (s)    & 5000 & 20000& 11000 & 16000            \\
$Q_{1}$ (erg/s)        & $5.0 \times 10^{47}$            & $5.0 \times 10^{47}$
                       & $8.0 \times 10^{47}$            & $9.0 \times 10^{47}$ \\
$t_{\rm start,2}$ (s)  &      &      &     & 65000            \\
$t_{\rm end,2}$ (s)    &      &      &     & 81000             \\
$Q_{2}$ (erg/s)        &             &
                       &             & $6.0 \times 10^{46}$\\
\enddata
\end{deluxetable}

\section{Discussion and Conclusions}
\label{sect:disc}

The progenitors of long GRBs may be WR stars (Chevalier \& Li 1999) within the context of the collapsar model
(Woosley 1993). Thanks to the localization ability of the $Swift$ satellite, many unexpected behaviors have been
observed in early GRB afterglows, such as significant optical rebrightenings. Such observations challenge the view
that afterglow light curves in the optical band should be smooth (Laursen \& Stanek 2003), which was the accepted
dogma since the discovery of the first GRB afterglow (e.g. Sahu et al. 1997).

In this paper, based on the fall back accretion model, we present a numerical study of the early X-ray afterglow emission
of GRBs 080810, 081028 and 091029, and compare our results with the observations. It is shown the X-ray bumps in the
early stage and the late X-ray plateaus or rebrightenings can be satisfactorily reproduced. We also apply our model
to the ultra-long GRB 111209A, which is characterized by complex afterglow light curves with an unprecedented burst
duration of a few hours (Hoversten et al. 2011). Distinguishing features of this event include the significant bump
at X-ray wavelength in the early stage and the remarkable optical rebrightening observed at about $10^{5}$ s after
the $Swift$/BAT trigger. We show that the early bump can also be fitted well by the fall back accretion model with
appropriate choices of parameters. For the late time afterglow, the significant optical rebrightening observed at
about $10^{5}$ s can be explained well with a constant energy injection process.

Another distinguishing feature of GRB 111209A is the long duration of the prompt emission, which is at least
25000 s as estimated by Gendre et al. (2013), based on the start time of the detection by KW and the start time
of the sharp decay. Assuming that the central engine can be kept active while the progenitor envelope
is being accreted onto it (Kumar, Narayan \& Johnson 2008a), the fall back accretion of a progenitor envelope (Quataert \&
Kasen 2012; Wu et al. 2013) may be one reasonable explanation for the long prompt emission duration of GRB
111209A. Woosley \& Heger (2012) pointed out that direct envelope collapse of a massive star can generate GRBs
with prompt emission duration of about $10^{4}$ s, therefore, a massive and extended progenitor like blue
supergiant (${\rm M\acute{e}sz\acute{a}ros}$ \& Rees 2001) is another possible choice as the central engine
of GRB 111209A. Gendre et al. (2013) argued that the collapse of a blue supergiant is the best candidate for
the progenitor of the ultra-long GRB 111209A, which was revised by Nakauchi et al. (2013) to explain the
durations of ultra-long GRBs, such as GRBs 111209A, 101225A, and 121027A. Actually, considering the X-ray bump
observed in the early stage and the ultra-long prompt emission duration, the fall back accretion is a
more convincing model for GRB 111209A. Detailed theoretical models of the ultra-long GRB 111209A were
discussed in Gendre et al. (2013). Interestingly, Zhang et al. (2013) found GRB 111209A is not so
unique by proposing a new definition of the duration of GRB to measure the true time scale of the central
engine activity from the physical point of view.

To reproduce the early X-ray bumps of the above GRBs, we invoke the BZ mechanism by assuming the central
compact object is a BH. Yoon, Langer \& Norman (2006) investigated the final fate of massive stars as a function of initial
mass and spin rate, at four different metallicities. In the GRB production region with low metallicity, BHs are expected
to form inside WR stars, where the minimum mass for BH formation is simply assumed to be 30 $M_{\odot}$. For the BZ
mechanism to work, there are many necessary requirements, among which whether the fall-back material is able to circularize
and form an accretion disk or not is an important one. This because is the magnetic field of the BH cannot be supported by the BH alone,
but can be assumed to be supported by a magnetized accretion disk (Lee et al. 2000). Perna et al. (2014) performed a numerical investigation of the
formation of fall back disks during a supernova explosion and found that an extended, long-lived disk can form in a
large region of the parameter space, largely independent of the role played by the magnetic torques in the evolution of the massive star.

Internal extinction in the host galaxies of GRBs has been considered by many authors in
explaining some GRBs, such as GRB 970508 (Sokolov et al 2001), GRB 980329 (Draine 2000), GRB 980613
(Sokolov et al 2001), GRB 051022 (Rol et al 2007) and GRB 081029 (Yu \& Huang 2013).
We have also taken extinction into consideration in our fit to the afterglow light curve of GRB 111209A through a correction
of 0.7 mag in the optical band. This agrees with the result derived by Stratta et al. (2013) who found a visual dust extinction
in the rest frame of order $A_{V} = 0.3 - 1.5$ mag during the prompt phase. This may be one possible reason for the non-detection
of the accompanied supernova of GRB 111209A. Interestingly, according to Stratta et al. (2013), there is no evidence of dust
extinction in the optical band after the steep decay phase.

In our calculations, we assumed a constant energy injection to interpret the X-ray plateau or rebrightening
observed at late times. As discussed by Yu \& Huang (2013), physically, the injected energy can be generated by the infall
of materials onto the central compact object of the burst. The envelope material that have survived mass loss will fall back if
its kinetic energy is less than the potential energy due to the central remnant. The fall back is usually continuous, but clumps could
sometimes exist in the fall back material. The fall back accretion rate will be significantly increased when a large clump suddenly
plunges towards the central BH. Through the BZ mechanism, the rotational energy of the black hole with an external
magnetic field supported by a surrounding disk can be extracted as a Poynting flux, which will eventually catch up with the outflow
and inject the energy into the external shock, producing a plateau or a rebrightening in the afterglow light curve.

In our current study the goodness of fit is not judged by a precise statistical criterion, so the error bars of the
parameters are not given. It should be noted that, a more precise fitting is actually not more informative because there
are many parameters involved in our numerical calculations. However, the main features of the light curves have been reproduced
by our fitting and our main purpose is to show that the fall back accretion process and the BZ mechanism are possible
in GRBs and a group of events may actually be generated in this way.

To conclude, it is shown that our model is consistent with the observations of these four GRBs.
The observed X-ray bump in the early stage can be reproduced by the fall back process and the X-ray plateau or rebrightening
at late times can be fitted very well with a constant energy injection. It is argued that the energy injection process can be
produced by materials that fall back onto the central compact object, which leads the accretion rate to increase and results
in a strong Poynting-flux-dominated outflow. In the future, more detailed studies on the energy injection and fall back processes
should provide important clues on the progenitors of GRBs, especially GRBs that show exceptional temporal features such as GRB 111209A.

\acknowledgments

\appendix
This work was supported by the National Basic Research Program of China (973 Program, Grant No. 2014CB845800) and the National Natural
Science Foundation of China (Grant Nos. 11473012, 11033002, 11322328 and J1210039). X.F. Wu acknowledges support by the One-Hundred-Talent Program,
the Youth Innovation Promotion Association, and the Strategic Priority Research Program ``The Emergence of Cosmological Structures''
(Grant No. XDB09000000) of Chinese Academy of Sciences. D.M. Coward is supported by an Australian Research Council Future Fellowship.
E.J. Howell acknowledges support from a UWA Research Fellowship.

\end{document}